\documentclass[11pt]{article}
\usepackage{moriond,psfig}

\def\Journal#1#2#3#4{{#1} {\bf #2}, #3 (#4)}

\def\be{\begin{equation}}
\def\ee{\end{equation}}
\def\bea{\begin{eqnarray}}
\def\eea{\end{eqnarray}}

\def\lsim{\lower 2pt \hbox{$\, \buildrel {\scriptstyle <}\over
         {\scriptstyle \sim}\,$}}
\def\gsim{\lower 2pt \hbox{$\, \buildrel {\scriptstyle >}\over
         {\scriptstyle \sim}\,$}}
\begin{document}
\vspace*{4cm}
\title{GAMMA RAYS FROM ROTATION-POWERED PULSARS}

\author{ALICE K. HARDING}

\address{Code 661, NASA Goddard Space Flight Center, Greenbelt, MD 20771}

\maketitle\abstracts{
The seven known gamma-ray 
pulsars represent a very small fraction of the more than 1000 presently known radio 
pulsars, yet they can give us valuable information  about pulsar particle acceleration 
and energetics.  Although the theory of acceleration and high-energy emission in 
pulsars has been studied for over 25 years, the origin of the pulsed gamma rays is 
a question that remains unanswered. Characteristics of the pulsars detected by 
the Compton Gamma-Ray Observatory 
could not clearly distinguish between an emission site at the magnetic poles (polar 
cap models) and emission from the outer magnetosphere (outer gap models).  There are 
also a number of theoretical issues in both type of model, which have yet to be resolved.  
The two types of models make contrasting predictions for the numbers of radio-loud 
and radio-quiet gamma-ray pulsars and of their spectral characteristics.  GLAST will 
probably detect at least 50 radio-selected pulsars and possibly many more radio-quiet 
pulsars.  With this large sample, it will be possible to fully test the model 
predictions and finally resolve this longstanding question.}

\section{Introduction}

The Compton Gamma-Ray Observatory (CGRO), which
ended its spectacularly successful mission in June 2000, provided a
wealth of new data on isolated, rotation powered pulsars.  While that data went 
a long way in constraining emission models, it was not enough to define the location
or mechanisms of the high energy pulsed emission and ultimately, of the particle
acceleration.  Meanwhile, as we have been waiting for the launch of the next $\gamma$-ray 
telescopes, Integral, AGILE and GLAST, there have been a large number of new pulsars detected
at radio and X-ray wavelengths by the Parkes Multibeam survey \cite{Man01}
and by the Chandra X-ray Observatory.  Many of these are young and energetic pulsars,
prime $\gamma$-ray pulsar candidates and many of them lie in or near the error boxes of 
unidentified $\gamma$-ray sources \cite{Gren02}.  Observations of these sources at
$\gamma$-ray wavelengths promise a harvest of new data for the study of high-energy
emission from isolated pulsars.  
   
\section{Summary of Observations}

During its nine-year mission, observations by CGRO increased the number of known
$\gamma$-ray pulsars from two to seven.  In addition, EGRET made four ``candidate"
detections with somewhat lower significance, including the millisecond pulsar 
PSR J0218+4232  \cite{Kui00}.  Table 1 lists the known $\gamma$-ray pulsars together
with the rotation-powered X-ray pulsars.  There are three main groups of sources in 
Table 1: pulsars detected only in X rays, those detected in X rays and radio and
those detected in $\gamma$ rays, X rays and radio.  One interesting and important
question is the number of high-energy pulsars that are radio quiet.  Since it is
generally believed that radio emission originates near the polar caps of the neutron 
star, lack of detection in the radio (if due to unfavorable beaming) would
strongly constrain the location and geometry of the high-energy emission.  All of the known
$\gamma$-ray pulsars are radio-loud with the possible exception of Geminga, which 
has unconfirmed reports of weak radio emission \cite{MM98}.  But they are
also radio
selected, since EGRET was not capable of independent period searches.  The ten 
radio-loud X-ray pulsars which were not detected as $\gamma$-ray pulsars fall into 
several categories.  The first five are millisecond pulsars (only one $\gamma$-ray
pulsar is a ms pulsar), the last three are young pulsars in supernova remnants,
too recently discovered to have provided concurrent ephemerides for CGRO detections,
PSR B1929+10 is a $10^7$ yr old pulsar with only thermal X-ray emission and PSR B0540-69 
is a young pulsar in the LMC.  Finally, the seven pulsars detected only in X rays are
a collection of all categories and may be truly radio quiet.  However, it has recently
become apparent that lack of detection by even the most sensitive radio surveys is
not proof that the pulsar is radio quiet.  J11245916 \cite{Cam02a,Hughes02} and 
J0205+6449 \cite{Mur02,Cam02b} were recently discovered to be very weak radio pulsars 
only after long, pointed observations.  The unexpected finding that these young 
pulsars have 
extremely low radio luminosities is raising new questions about the radio luminosity
function and its implications for the neutron star birthrate.

\begin{table}[t]
\caption{\large High-Energy Pulsars}
\vspace{0.4cm}
\begin{center}
\begin{tabular}{|l|c|c|c|c|c|l|}
\hline						
&&&&&& \\										
PSR &	P (ms)  &  $\dot P$ ($10^{-15}\rm s\,s^{-1}$) & Gamma-ray	& X-Ray &	Radio & SNR/Name	\\			
\hline										
B0531+21 &	33.0 & 422.0 &    X &	X &	X &	Crab	\\			
B1509-58 &	150.0	& 1540.0 &	X &	X &	X &	MSH15-52 \\				
B0833-45 &	89.0 & 124.0 &    X &	X &	X &	Vela	\\			
B1706-44 &	102.0 & 92.2 &    X &	X &	X &	\\				
B1951+32 &	39.5 & 5.9	&     X &	X &	X &	CTB 80 \\				
J0633+1746 & 237.0 & 11.4 &	X &	X &	(X) &   Geminga \\				
B1055-52 &	197.0 & 5.8 &     X &	X &	X &	\\				
B0656+14 &	384.0 & 55.0 &	(X) &	X &	X &	\\				
J0218+4232 & 2.3 & 8.3E-05 &	(X) &	X &	X &	\\				
B1046-58 &	124.0	& 103.6 &	(X) &	X &	X &	\\				
J2229+6114 & 51.6	& 78.3 &	(X) &	X &	X &	\\				
\hline 										
B1957+21 &	1.60 & 1.1E-04	&&	X &	X & \\					
J0437+4715 & 5.70 & 5.7E-05	&&	X &	X & \\					
J0030+0451	&4.86	& 1.0E-05	&&	X &	X & \\					
J2124-3358	&4.93	& 2.1E-05	&&	X &	X & \\					
B1821-24 &	3.05	& 1.6E-03	&&	X &	X & \\					
B1929+10 &	226.0	& 1.2 	&&	X &	X & \\					
B0540-69 &	50.0	& 479.0	&&	X &	X &	 \\				
J1420-6048 & 68.0 & 83.0	&&	X &	X &	Kookaburra \\
J1124-5916 & 136.0 & 740.0	&&	X &	X &	G292.0+1.8 \\			
J0205+6449 & 65.7	& 193.0	&&	X &	X &	3C 58 \\			
\hline 										
J0537-6910 & 16.1 & 51.2 	&&	X &&        N147B \\						
J0635+0533 & 33.9	& *		&&    X &&	\\								
J1811-1926 & 64.7	& 44.0	&&	X &&	      G11.2-0.3 \\					
J1846-0258 & 324.0 &	7097.1&&	X &&		Kes75	\\			
J161730-50 &  69.0 &	140.0	&&	X &&		RCW 103 \\	
J0822-43   & 75.3	&149.0	&&	X &&		Puppis A \\
1E1207.4-52 & 424.0&		&&	X &&		PKS1209-52	\\
\hline 
\end{tabular}
\end{center}
\end{table}

Figure 1 plots the known radio and $\gamma$-ray pulsars as a function of their period, 
$P$, and period derivative, $\dot P$.  The $\gamma$-ray pulsars reside in the upper-left
corner of the radio pulsar population, and with the exception of the millisecond
pulsar J0218+4232, all have ages less than $10^6$ yr.  Also shown are the sixteen radio 
pulsars associated with EGRET unidentified sources.  It is remarkable that with one 
exception, all have ages less than $10^5$ yr, suggesting that some real associations of
the $\gamma$-ray emission with emission from the pulsars or their nebulae is likely.

A pattern which emerged from the $\gamma$-ray pulsar light curves measured by CGRO was
that, with the one exception of PSR1509-58, all have double-peaked pulses with 
interpeak emission.  This strongly suggests that we are observing emission 
associated with just one of the magnetic poles and that the emission pattern is a 
hollow cone or wide fan beam.  Most of the pulsars detected only at X-ray energies,
by contrast, have broad single-peaked pulses.  In most cases, the radio pulses lead the
$\gamma$-ray pulses in phase rather than being coincident.
Another strong pattern identified in the CGRO pulsars 
was a correlation between
high-energy luminosity and the quantity $P^{-3/2}\dot P^{1/2}$, where $P$ and $\dot P$
are the pulsar rotation period and period derivative, respectively.  For a pure dipole
field, this quantity is proportional to the polar cap current, the voltage across open 
field lines and $\dot E_{\rm rot}^{1/2}$, where $\dot E_{\rm rot}$ 
is the spin-down luminosity.  Thus
the $\gamma$-ray luminosity seems to be closely tied to the primary particle
acceleration of the pulsar. 

\begin{figure}[t]
\vskip -0.7in
\hskip 1.0in
\psfig{figure=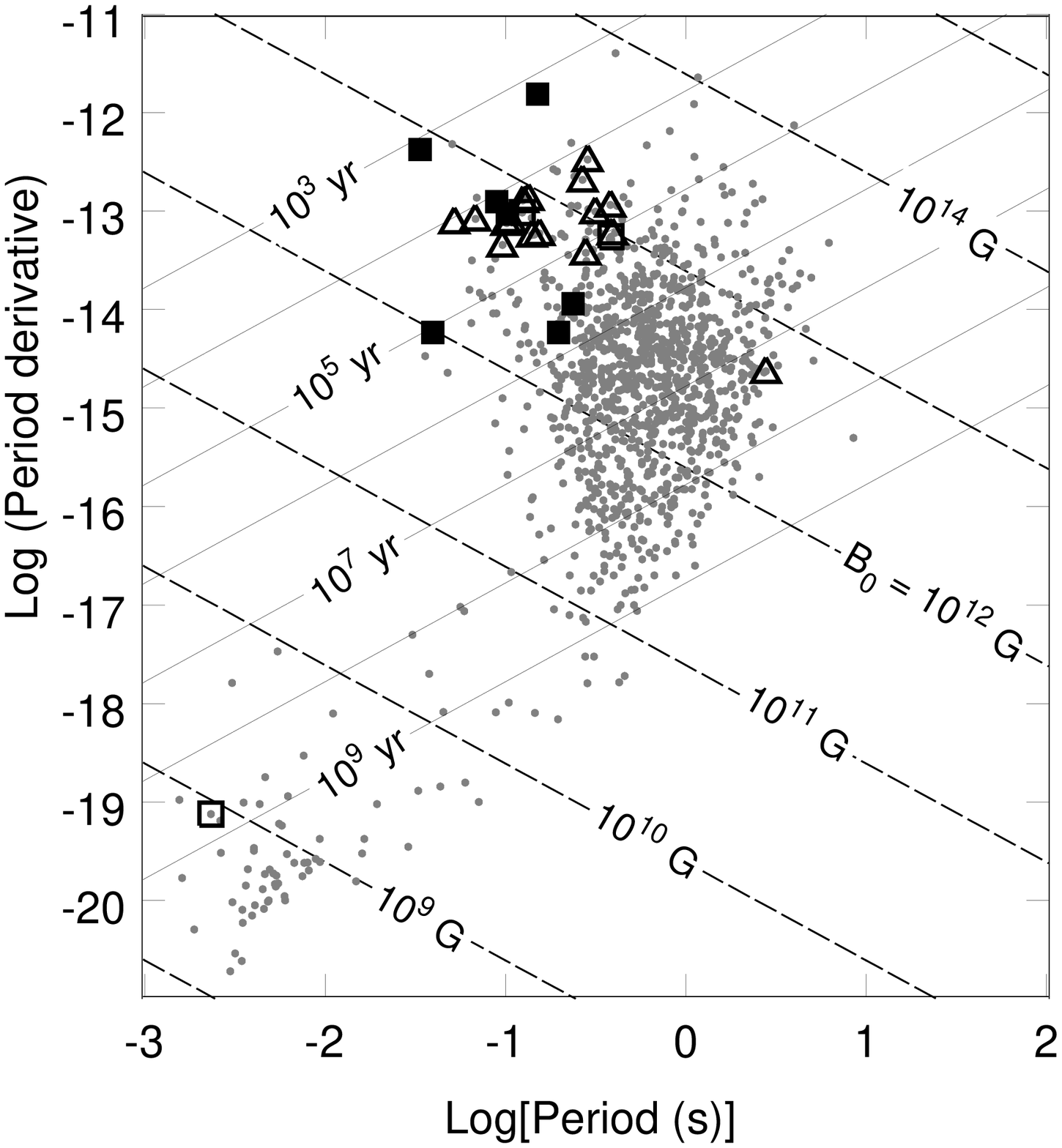,height=6in}
\caption{Plot of period-derivative $\dot P$ vs. period $P$ of radio pulsars (gray 
dots) from the ATNF Pulsar Catalog (http://www.atnf.csiro.au/pulsar/) with confirmed 
(solid squares) and candidate (open squares) $\gamma$-ray 
pulsars.  Also plotted are radio pulsars in EGRET unidentified source
error boxes (open triangles).  Solid diagonal lines indicate spin-down age, $\tau = 
P/2\dot P$.
\label{fig1}}
\end{figure}

The broad-band spectra from radio to TeV energies of the seven confirmed CGRO 
$\gamma$-ray pulsars \cite{Thomp97} show that the emission power in these sources
peaks in the hard X-ray or $\gamma$-ray part of the spectrum. 
The spectra of all the CGRO pulsars have clear high-energy turnovers since none have
detected pulsed emission above 20 GeV.  Three pulsars, Vela, Crab and Geminga, have 
spectral turnovers
in the EGRET range, around 5 GeV and one, PSR1509-58, has a sharp turnover in the
COMPTEL range around 10 MeV.  Below the turnovers, there is a trend of increasing
spectral hardness with dipole spin-down age, $\tau = P/2\dot P$. 
In the middle-aged and older pulsars, a thermal component in the soft X-ray band 
is visible above the falling power-law continuum.  This component is probably present 
also in young pulsars but hidden beneath the strong power-law.

\section{Emission Models and Predictions}

Particle acceleration inside the pulsar magnetosphere gives rise to pulsed 
non-thermal radiation. 
Rotating, magnetized neutron stars are natural unipolar inductors, 
generating huge ${\bf v x B}$ electric fields.  However, these fields 
are capable of pulling charges out of the star 
against the force of gravity and it is believed that the
resulting charge density that builds up in a neutron star magnetosphere
is able to short out the electric field parallel to the magnetic field 
($E_{\parallel}$)(thus allowing the field to corotate with the star) 
everywhere except at a few locations.  Two possible sites where 
${\bf E \cdot B} \neq 0$, and strong $E_{\parallel}$ may develop to accelerate
particles, have given rise to two classes of high energy emission models: polar 
cap models, where the acceleration and radiation occur at the magnetic poles close 
to the neutron star surface, and outer gap models, where these processes occur in 
the outer magnetosphere.

\subsection{Polar cap}

Polar cap models  \cite{DH96,UM95} 
advocate that particle acceleration occurs near the neutron star surface and that
$\gamma$-rays result from a curvature radiation or inverse Compton induced 
pair cascade in a strong magnetic field. There is some variation among 
polar cap models, with 
the primary division being whether or not there is free emission of particles 
from the neutron star surface. The subclass of polar cap models based on free 
emission of particles of either sign, called space-charge limited flow models, 
assumes that the surface temperature 
of the neutron star (many of which have now been measured in the range 
$T \sim 10^5 - 10^6$ K) exceeds the ion and electron
thermal emission temperatures.  Although $E_{\parallel} = 0$ at the neutron star 
surface in these models, the space charge along open field lines above the surface
falls short of the corotation charge, due to the curvature of the field  \cite{Arons83}
or to general relativistic inertial frame dragging  \cite{MT92}.
The $E_{\parallel}$ generated by the charge deficit accelerates particles, which
radiate inverse Compton (IC) photons (at particle Lorentz factors 
$\gamma \sim 10^2 - 10^6$) and curvature (CR) photons (at Lorentz factors 
$\gamma \gsim 10^6$).  Recent studies  \cite{HA01,HM02a} have 
found that virtually {\it all} pulsars can produce IC radiation, by scattering thermal
X rays from the NS surface, at high enough energies to produce pairs.  The IC emission 
is mostly resonant scattering in high-field pulsars and non-resonant scattering
in older and weaker field pulsars.  Because lower Lorentz factors are 
required to produce pairs
through IC emission, an IC pair formation front (PFF) will form first, close to
the surface.  However, it is found  \cite{HM02a} 
that the IC pair formation fronts do not produce sufficient pairs to 
screen the $E_{\parallel}$ completely, thus allowing acceleration to the Lorentz
factors $\sim 10^7$ sufficient to produce a CR PFF, where there are sufficient pairs
to completely screen the $E_{\parallel}$.  The CR pair front will therefore limit
the particle acceleration voltage and determine the high-energy emission luminosity.

It has been realized for some time that only younger pulsars
are capable of producing pairs (and pair cascades) through CR emission in dipole
magnetic fields  \cite{Arons83}.  These pulsars with 
\be
{\dot E}_{\rm rot} > {\dot E}_{\rm rot, break} = 5 \times 10^{33}~P^{-1/2}
~~~{\rm erg}\cdot {\rm s}^{-1}.
\ee
will have screened $E_{\parallel}$, so that the acceleration voltage $\Phi$ is nearly 
independent \cite{HM02a,HM02b} of the period and surface magnetic field, $B_0$.  Pulsars 
having spin-down luminosity ${\dot E}_{\rm rot} < {\dot E}_{\rm rot, break}$ 
will not produce CR pairs to
screen $E_{\parallel}$, so that the acceleration voltage will be limited only by the
saturation due to the geometry of the polar cap and this $does$ depend on $P$ and $B_0$.  
The bolometric luminosity is $L_{\rm CR} \simeq \epsilon\,\Phi \,\dot n_{prim} \simeq 
\epsilon \Phi \,n_{GJ}\,\pi \,R_{PC}^2 c$, where $\dot n_{prim} 
\propto {\dot E}_{\rm rot}^{1/2}$ 
is the primary PC current, $n_{GJ}$ is the Goldreich-Julian 
density, $R_{PC}$ is the polar cap radius, and $\epsilon$ is a radiation effiency.  
We thus have the luminosity prediction 
\be \label{L}
L_{\rm CR} = 
\left\{ \begin{array}{lr}
\epsilon \,10^{16}~{\rm (erg/s)^{1/2}}~~\dot E_{\rm rot}^{1/2}
    P^{1/14}\,B_{12}^{-1/7} & \dot E_{\rm rot} > {\dot E}_{\rm rot, break} \\
\epsilon \,{3\over 4} \kappa (1-\kappa ) {\dot E}_{\rm rot} & \dot E_{\rm rot} < 
{\dot E}_{\rm rot, break}
\end{array}
\right. 
\ee
for PC space-charge limited flow acceleration \cite{HM02b}, where 
$\kappa = r_g I/MR^3 \simeq 0.14$, 
and $r_g$ is the gravitational radius of a NS of mass $M$, and $I$ and $R$ are the stellar 
moment of inertia and radius, respectively.  The above expression is valid for surface
fields $B_0 \lsim 5 \times 10^{12}$ G.
Figure 2 shows the predicted luminosity from Eqn (\ref{L}), assuming $\epsilon = 0.5$, and
the luminosities of the observed $\gamma$-ray pulsars.  The break from $L_{\rm CR} 
\propto 
{\dot E}_{rot}$ to $L_{\rm CR} \propto {\dot E}_{rot}^{1/2}$ is predicted to occur for
luminosities just below that of the detected $\gamma$-ray pulsars.

\begin{figure}[t]
\vskip -0.5in
\hskip 0.5in
\psfig{figure=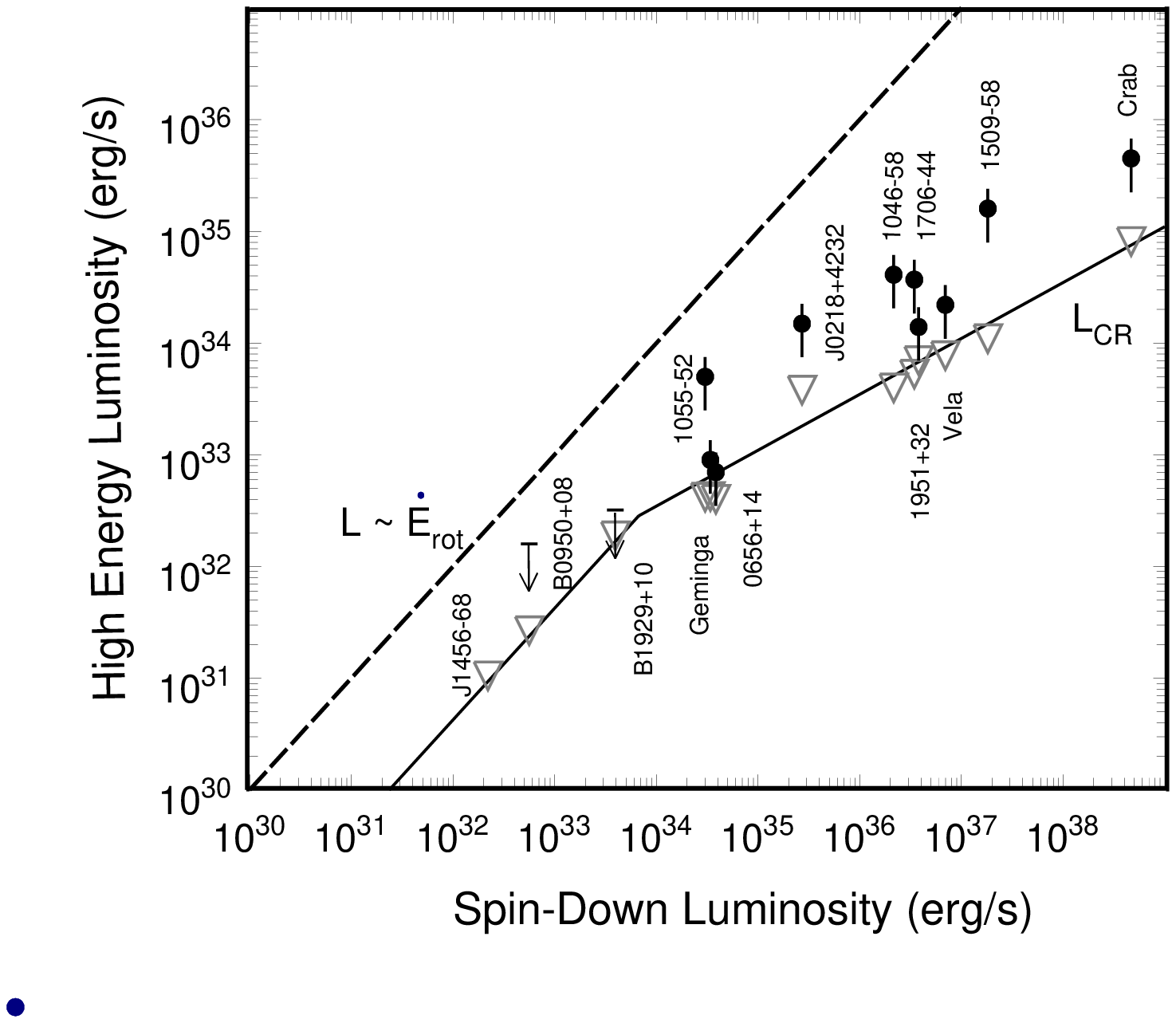,height=4.5in}
\caption{Predicted and observed high energy luminosity vs. spin-down luminosity.  The solid
curve is the theoretical prediction from eq. (\ref{L}).  
The solid circles are the luminosities of the detected $\gamma$-ray 
pulsars\protect \cite{Thomp01}, 
derived from detected fluxes above 1 eV assuming a 1 sr. solid angle. 
The upper limits 
are for $> 100$ MeV from\protect \cite{Thomp94}.  The open triangles are predicted
luminosities for the detected pulsars.}  
\label{fig2}
\end{figure}

Polar cap models predict that the $\gamma$-ray spectra cut off very sharply (as
a ``super-exponential") due to one-photon pair production attenuation, at the 
pair escape energy \cite{HBG97}, i.e. the highest energy at 
which photons
emitted at a given location can escape the magnetosphere without pair producing.
A rough estimate of this cutoff energy, assuming emission
along the polar cap outer rim, $\theta_{_{\rm PC}} \simeq (2\pi r / cP)^{1/2}$, 
at radius $r$, is \be \label{Ec}
E_c \sim 2\,\,{\rm GeV}\,P^{1/2}\,\left({r\over R}\right)^{1/2}\, {\rm max}
\left\{0.1, \,B_{0,12}^{-1}\,\left({r\over R}\right)^3\right\}
\ee
where $B_{0,12}$ is the surface 
magnetic field in units of $10^{12}$ G.  At all but the highest fields
there is a prediction that the spectral cutoff energy should be inversely
proportional to surface field strength, or $B_0 = 6.4 \times 10^7\,{P\dot P}^{1/2}$
G for a dipole field.  In fields above $\sim 2 \times 10^{13}$ G, photon splitting, 
in which a single photon splits into two lower energy photons, becomes the
dominant attenuation process and lowers the photon escape energy  \cite{HBG97,BH01}. 
The observed cutoff energies of eight $\gamma$-ray pulsars seem to
increase with decreasing surface field  \cite{Hard01a}, although a larger number of 
measurements are needed to confirm this trend.  

According to Eqn (\ref{L}), millisecond pulsars should have detectable high-energy
emission.  However only one, PSR J0218+4232, has been detected in $\gamma$ rays by
EGRET \cite{Kui00}.  But the polar cap particle acceleration in millisecond
pulsars will be limited by curvature radiation reaction \cite{Luo00,HM02b}
so that the CR emission spectrum will be quite hard (photon index $-2/3$).  
Because ms pulsars have surface fields of only $B_0 \sim 10^8 - 10^{10}$ G, the 
spectral cutoff will not be caused by magnetic pair production at an energy of a 
few GeV, but will occur at the cutoff of the CR spectrum, causing the $\nu F_{\nu }$ 
spectrum to peak at 50-100 GeV energies \cite{Bulik00}.  There may be a second
spectral component at X-ray energies from IC pair cascades.  This may explain why 
many millisecond pulsars detected at X-ray energies are not detectable by EGRET, but
may be visible to air Cherenkov telescopes having thresholds below 100 GeV.

\subsection{Outer gap}

Outer-gap models  \cite{CHR86,Rom96} assume that acceleration
occurs in vacuum gaps along null charge surfaces in the outer magnetosphere and that 
$\gamma$ rays result from photon-photon pair production-induced cascades.
The gaps arise because charges escaping through the light cylinder along open field
lines above the null charge surface cannot be replenished from below.  Pairs from the polar
cap cascades, which flow out along all the open field lines, will undoubtedly
pollute the outer gaps to some extent (and vice versa), 
but this effect has yet to be investigated.
The electron-positron pairs needed to provide the current in the outer gaps are 
produced by photon-photon pair production.
In young Crab-like pulsars, the pairs are produced by CR photons from the primary 
particles interacting with non-thermal synchrotron X-rays from the same pairs.
In older Vela-like pulsars, where non-thermal X-ray emission is much lower, the pairs
are assumed to come from interaction of primary particles with thermal X-rays from the neutron 
star surface.  Some of the accelerated pairs flow downward to heat the 
surface and maintain the required thermal X-ray emission.  The modern outer gap 
Vela-type models  \cite{Rom96,ZC97} all adopt this picture.  Although there
seems to be agreement on the radiation processes involved in the outer gap,
the full geometry of the gap is still not solved.  Two approaches to such a solution
are currently underway, but neither is near to defining the complete three-dimensional
gap geometry.  One group \cite{Rom96,CRZ00} solves the 1D 
Poisson equation perpendicular to the magnetic field lines, resulting in a gap geometry 
for young pulsars that is a long, thin sheet bounded by the last open field line. 
The other group~\cite{Hir00} obtains solutions to the 1D Poisson equation 
along the magnetic field (assuming a gap width across field lines) and finds that
the gap is limited parallel to the field by pair creation.  The actually gap geometry is
probably somewhere in between (see Figure 3).  The long narrow outer gap geometry of 
Yadigaroglu \& Romani \cite{YR95} and Cheng et al. \cite{CRZ00}, 
which has been so successful in reproducing the 
observed double-peaked pulse profiles needs to be re-examined to also explore gap
closure along the field.     

When the high-energy photons are emitted in the outer
magnetosphere, where the local magnetic field is orders of magnitude lower than
the surface field, one-photon pair production plays no role in either the pair cascade or
the spectral attenuation.  In this case the high-energy cutoffs in the photon spectrum 
come from the upper limit of the accelerated particle spectrum, due to radiation reaction.
The shape of the cutoff is thus a simple exponential, more gradual than in polar
cap model spectra.  Due to the large errors of the EGRET data points above
1 GeV, the measurements at present do not definitely discriminate between model
spectra.  GLAST should have the energy resolution and dynamic range to measure the shape 
of the cutoffs seen by EGRET and should be able to rule out either the simple exponential 
or super-exponential shape.  In addition, GLAST will detect enough $\gamma$-ray 
pulsars with different field strengths to look for a correlation between surface
field strength and cutoff energy.  

Outer gap models predict an emission component at TeV energies due to inverse Compton 
scattering by gap-accelerated particles.  TeV photons should escape from the outer gaps
of pulsars with even strong surface fields and this component is thus expected to be
observable in many pulsars.  The original predictions of Cheng et al.  \cite{CHR86} 
were not verified by observations of ground-based detectors  \cite{Nel93,Less00}, 
requiring a
revision of the Vela-like model  \cite{Cheng94}.  However, even later models which
predicted lower TeV fluxes  \cite{Rom96} are above CANGAROO upper limits on pulsed
emission from Vela  \cite{Yosh97}.  The most recent outer gap models 
 \cite{Hir00,HS01}, have predicted
TeV inverse-Compton fluxes which are below the present observational upper limits,
but which should be detectable with the next generation of TeV detectors.  Unfortunately,
while a TeV emission component is an essential prediction of all outer gap models,
the inverse Compton flux level depends on the pulsed emission spectrum in the 
infra-red (IR) band which is notoriously difficult to measure in most pulsars.

Unlike in polar cap models, pair production in outer gap models is essential to the
production of the high energy emission: it allows the current to flow and 
particle acccleration to take place in the gap. Beyond a death line in 
period-magnetic field space, and well before the traditional radio-pulsar death line, 
pairs cannot close the outer gap and the pulsar cannot 
emit high energy radiation.  This outer gap death line for $\gamma$-ray 
pulsars  \cite{CR93} falls around $P = 0.3$ s for $B \sim 10^{12}$ G.  Geminga
is very close to the outer gap death line and recent self-consistent models
 \cite{HS01} have difficulty accounting for GeV $\gamma$-rays from
pulsars of this age. Polar cap models, on the other hand, predict that all pulsars 
are capable of $\gamma$-ray emission at some level, so that detection as a radio-loud 
$\gamma$-ray 
pulsar is thus a matter of sensitivity.  Detection of pulsars with periods much 
exceeding that of Geminga would thus be strong evidence in favor of polar cap models.

\section{Geometry, Population Statistics and Radio-Quiet Pulsars}

\begin{figure}[t]
\psfig{figure=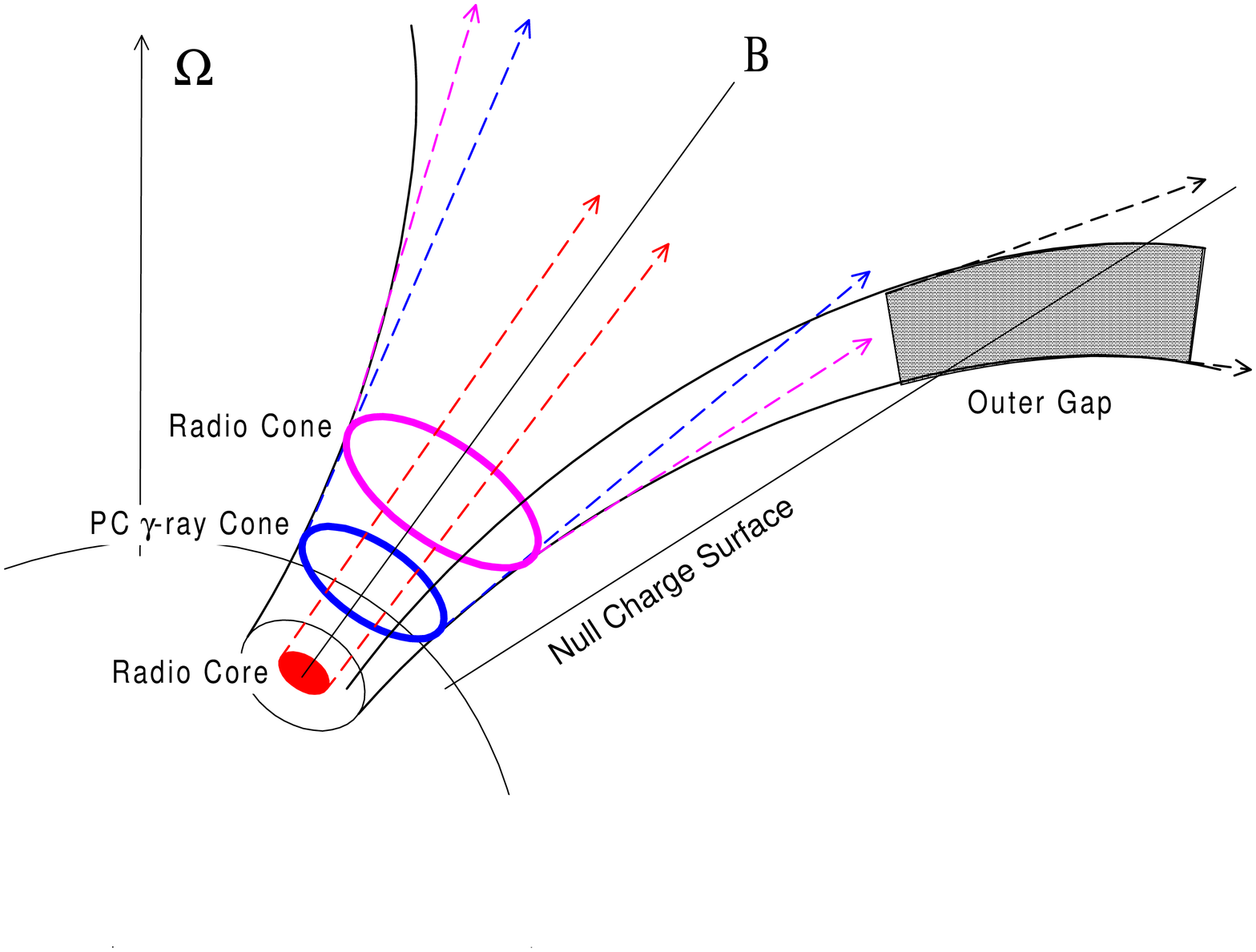,height=4.5in}
\vskip -0.5in
\caption{Cartoon of polar cap, outer gap and radio beams (not necessarily to scale).  
The outer gap is shown as a 
2D projection, so that the full 3D extent of the gap and beam is not represented.
\label{fig3}}
\end{figure}

In polar cap models, the $\gamma$-ray and radio emission are physically connected,
since the electron-positron pairs from the PC cascades are thought to be a necessary 
ingredient for coherent radio radiation.  However, the relative geometry of the two
emission regions is not very constrained from observation.  A possible scheme for the
geometry of the radio and high-energy emission beams is shown in Figure 3.  Models for
radio emission morphology \cite{Ran93} \cite{LM88} include core and conal 
components, emitted within a radius of about ten stellar radii from the surface. 
If the PC $\gamma$-ray cone emission occurs at a lower altitude than the radio emission,
as would be expected, then the leading edge of the radio cone would lead the first 
$\gamma$-ray peak.  The trailing edge of the radio conal emission would then have to be
undetected in the CGRO pulsars.  This type of radio geometry for young pulsars was
suggested by Manchester \cite{Man96} and there is indeed a class of pulsars with one-sided
radio conal emission \cite{LM88}.  Another requirement on the 
geometry of polar cap $\gamma$-ray
emission is that the inclination angle be comparable to or smaller than the size of the 
$\gamma$-ray emission beam.  Since the high-energy emission beam is comparable to the 
angle of the open field lines, $\theta_{PC} \sim (\Omega r/c)^{1/2}$, this requirement
is very restrictive unless the emission radius, $r$, is at least several stellar
radii \cite{DH96}.   

High energy emission in the outer gap is generally radiated
in a different direction from the radio emission, which allows these models to 
account for the observed phase offsets of the radio and $\gamma$-ray pulses.  At the
same time, there will be fewer radio-$\gamma$-ray coincidences and thus a larger
number of radio-quiet $\gamma$-ray pulsars.  In Romani \& Yadigaroglu's  \cite{RY94}
geometrical outer gap model, the observed radio emission originates 
from the magnetic pole
opposite to the one connected to the visible outer gap.  Many observer lines-of-sight
miss the radio beam but intersect the outer-gap $\gamma$-ray beam, having a much
larger solid angle.  When the line-of-sight does intersect both, the radio pulse
leads the $\gamma$-ray pulse, as is observed in most $\gamma$-ray pulsars.

The emission geometry greatly influences the predicted numbers of radio-loud and
radio-quiet $\gamma$-ray pulsars.  Polar cap $\gamma$-ray emission is expected
to have a much higher observational correlation than outer-gap emission with the radio emission.
Simulations of the radio and $\gamma$-ray pulsar populations in both models
confirm this.  Although there
are significant variations in the numbers of predicted $\gamma$-ray pulsars 
due to different model assumptions in the various studies, outer gap models
clearly predict a much larger ratio of radio-quiet to radio-loud $\gamma$-ray
pulsars.  Polar cap model simulations \cite{Gon02} find that EGRET should
have detected 
many fewer radio-quiet than radio-loud pulsars,
while outer gap model simulations~\cite{YR95,ZZC00} claim that many of the 
40-60 EGRET unidentified sources at low latitudes are radio-quiet pulsars.
However, pulsars are ``radio-quiet" in PC models primarily because their flux is below the
sensitive limit of present radio surveys but in outer gap models primarily because the
radio emission is beamed away from us.     
The polar cap model simulations \cite{Gon02} predict that GLAST should detect slightly 
more radio-quiet than radio-loud pulsars as point sources.  Although GLAST will 
have the capability to detect pulsed $\gamma$-ray emission, the required sensitivity is 
much higher, about equal to the EGRET point source sensitivity. Thus, only about 10\% 
of the radio-quiet sources will have detected $\gamma$-ray pulsations.   
The outer gap simulations  \cite{ZZC00} predict that GLAST may detect 13 times
as many radio-quiet as radio-loud pulsars as point sources, with the detected 
number of radio-quiet pulsars equaling the present radio pulsar population!  This
would have profound consequences for neutron star evolution and supernova rates in
the galaxy.

Another possible population of radio-quiet $\gamma$-ray pulsars has been
suggested  \cite{HZ01}.  According to
the polar cap model \cite{DH96}, $\gamma$-ray emission occurs
throughout the entire pulse phase.  Primary electrons that initiate pair cascades at
low altitude continue to radiate curvature emission on open field lines to high 
altitudes beyond 
the cascade region, producing a lower level of softer off-beam emission.  Due to the 
flaring of the dipole field lines, this emission may be seen over a large solid angle, 
far exceeding that of the main $\gamma$-ray and radio beams.  It is therefore quite 
probable to see the off-beam $\gamma$-ray emission and miss the radio beam.  But since
it is much less luminous than the on-beam emission, off-beam emission will be detectable
only in nearby sources.  EGRET detected a population of unidentified $\gamma$-ray sources
correlated with the Gould Belt \cite{Gehrels00}, a young star cluster only $~100-300$
pc away, and recent simulations of pulsars in the Gould Belt \cite{GP01,Gren02} suggest that
such off-beam $\gamma$-ray emission might plausibly account for a significant fraction
of these sources. 

\section{Prospects for the Future}

Our window on the $\gamma$-ray universe from space will open again soon with the launch of
Integral~\cite{Her02} in October 2002, AGILE~\cite{Chen02} in 2003 and ultimately 
GLAST~\cite{Gren02} in 2006.  Searches for pulsed $\gamma$ rays in EGRET unidentified 
source error boxes with simultaneous timing of the new radio pulsars associated with
these sources
may finally provide some counterparts.  Searching for $\gamma$-ray emission from the
recently discovered X-ray pulsars inside young supernova remnants may also be fruitful.
Although it is only slightly more sensitive than EGRET, AGILE may significantly add to
the $\gamma$-ray pulsar population due to the tripling of the radio pulsar population
since the death of CGRO.  With its better sensitivity at energies above 1 GeV, AGILE
will be able to improve measurements of the spectral cutoffs detected by EGRET and
discriminate between emission models.  GLAST will detect at least an order of magnitude
more radio-loud pulsars and many more radio-quiet pulsars.  It will, for the first time,
be capable of doing blind period searches in about 10\% of the source population in
the plane and all of the source population detected by EGRET in the Gould Belt.
Meanwhile, more sensitive ground-based air Cherenkov telescope~\cite{Smith02} 
may be able to detect
the pulsed inverse-Compton emission around 1 TeV predicted by outer gap models or the
pulsed curvature radiation around 50-100 GeV from millisecond pulsars predicted by polar cap 
models.  Upper limits at this energy are also very important in constraining emission
models.  In summary, the next five years promise to be exciting ones for 
$\gamma$-ray astronomy.

\end{document}